\documentclass{article}

\usepackage{graphicx} 
\usepackage{url}

\title{Towards Properly Implementing Theory of Mind in AI: An Account of Four Misconceptions}
\author{Ramira van der Meulen, Rineke Verbrugge, Max van Duijn}
\date{February 2025}

\begin{document}

\maketitle


\noindent \textbf{This is the draft version of a paper that we expect to submit in April 2025. While the contents in the paper are not binding, this draft version should provide a fairly accurate depiction of the final contents of the paper. For questions about the contents of the paper, please contact the first author (A.van.der.Meulen@leidenuniv.liacs.nl).}

\section{Introduction}

The search for effective collaboration between humans and computer systems is one of the biggest challenges in Artificial Intelligence. One of the more effective mechanisms that humans use to coordinate with one another is theory of mind (ToM). ToM can be described as the ability to `take someone else's perspective and make estimations of their beliefs, desires and intentions, in order to make sense of their behaviour and attitudes towards the world'. If leveraged properly, this skill can be very useful in Human-AI collaboration.\\
\\
This introduces the question how we implement ToM when building an AI system. Humans and AI Systems work quite differently, and ToM is a multifaceted concept, each facet rooted in different research traditions across the cognitive and developmental sciences. We observe that researchers from artificial intelligence and the computing sciences (AI \& CS), ourselves included, often have difficulties finding their way in the ToM literature. In this paper, we identify four common misconceptions around ToM that we believe should be taken into account when developing an AI system. We have hyperbolised these misconceptions for the sake of the argument, but add nuance in their discussion. We end each section by providing tentative guidelines on how the misconception can be overcome.

\subsection{Introducing the Four Misconceptions}

The first misconception we discuss concerns modularity. Human ToM is likely the result of multiple brain processes working together, whereas in AI \& CS it is often conceptualised as a single module that adds a separate reasoning component to the system. In other words: (1) \textbf{``Humans Use a ToM Module, So AI Systems Should As Well''}.\\
\\
The second misconception concerns when to use ToM to begin with, as we sometimes fall into the trap of thinking that (2) \textbf{``Every Social Interaction Requires (Advanced) ToM''}. We know that adding a `stronger' reasoner to a system can improve its performance, but it is by far not always the more realistic option, also not when looking at the literature on human problem-solving. In fact: In everyday interaction, humans are more likely to use reasoning shortcuts than to overanalyse the situation. A system designed to understand human perspectives should take this into account.\\
\\
The third misconception concerns limited universality. ToM might be a universal skill among humans, but its expression has differences depending on the human in question (for example, neurological, but also cultural factors play a part). These differences in expression are even stronger when dealing with an entity that does not perceive nor process the world in the same way as a human (such as an AI system). It is important to realise, that in fact, not (3) \textbf{``All ToM is the Same''}.\\
\\
Finally, we address a recent popular claim. It is, for lack of a better term, often said that (4) \textbf{``Current Systems Already Have ToM''}. While some systems certainly are able to evaluate ToM on a level of specific tasks, recent models, especially large language models (LLMs), have evoked claims of generalisable ToM. These claims stem from their strong performance at many ToM benchmark tests. While not underestimating recent achievements, we nuance such claims by discussing generalisability of performance beyond linguistic platforms, keeping a clear view of the limitations and challenges ahead.

\section{Definitions} \label{definitions}

While a brief background to most of the concepts and terms is provided in the course of our discussion, there are a few definitions that we highlight beforehand. Apart from `Theory of Mind' and its relatives `Mindreading', `Perspective-Taking', and `Mentalisation', these are the `Beliefs, Desires, Intentions Architecture (BDI)', and a distinction between the `Behavioural' and `Mechanistic' level of analysis.\\
\\
\textbf{Theory of Mind (ToM)} The capacity central in this paper has its roots in debates in philosophy of mind \cite{dennett1971intentional} and ethology \cite{premack1978does} from the 1970s. It goes back to the basic observation that humans and some other animals, in particular great apes, are able to predict how another individual will act in the near future, and factor this into the planning of their own behaviour. In order to do this, they must reason from the other's perspective and thus have an understanding (a `theory') of the way they perceive and think about the world (cq. the other's `mind'). In the ensuing decades, ToM was studied extensively in psychology, where it was soon linked to atypical developmental patterns such as autism \cite{baron1985}, a view that was later challenged (see Section \ref{humanhuman} below). While various perspectives on and definitions of ToM have emerged (such as \cite{apperly2009humans,vanduijn2016,hutto2008folk,apperly2010mindreaders}, there are two core aspects that we also include in our definition here: ToM is i) making sense of behavioural elements that one \textit{can} observe by ii) reasoning in terms of what one \textit{cannot} observe: i.e. mental states including beliefs, desires, intentions, motivations, and so on. Such sense-making can serve the purpose of predicting what someone else will do, but also help to deepen one's understanding of the other's perspective without immediate behavioural consequences––or of oneself, given that ToM can also be applied self-reflectively. The terms \textbf{Mindreading}  and \textbf{Perspective-Taking} are used synonymously with ToM in this paper.\\
\\
\textbf{Mentalisation} One's internal picture of someone else's mental states, and why they act the way they do (also: to mentalise, the action of forming this picture). In other words: The result of having applied ToM to someone.\\
\\
\textbf{Beliefs, Desires, Intentions Architecture (BDI)} A popular AI model that can store and leverage the beliefs, desires and intentions of the entities it encounters \cite{rao1995bdi}. The model is occasionally also leveraged to represent additional mental states, such as explicit goals.\\
\\
\textbf{Behavioural vs. Mechanistic Level} We distinguish actions on a mechanistic and on a behavioural level. On a behavioural level, it is difficult to distinguish between ToM-style reasoning, and a pre-learned response (f.e. by an associative reasoner). Imagine you open a door for a colleague. Whether opening this door happens because you have explicitly taken their perspective and reasoned `this is what they want, and I would like to stay in their good graces', or because `it feels right, and is what I always do' is visibly indistinguishable to both the colleague and any outside observer. Mechanistically, they are vastly different, however. In the first case, the mechanism is `ToM'. In the second case, it's an experience-based heuristic.

\section{Humans Use a ToM Module, So AI Systems Should As Well}

For the discussion of our first misconception, suppose we are building a robot assistant in healthcare. It needs to be able to move around, move its arms, respond to patients, see the world, and so on. Suppose each of these functions is regulated through the robot's brain, its controller. We could be tempted to add a `module' that regulates (pro)social behaviour, enabling the assistant to understand and anticipate other entities (humans) around it, and call this its `ToM module'. This ToM module can be used for \emph{(1) Finding out a hospital patient's specific want, and asking relevant questions} (other-inquiry, empathy, thought evaluation), \emph{(2) realising that the patient moved to the right to get closer to an object they desire} (spatial reasoning, desire evaluation), \emph{(3) working with a patient to get them to take the right medicine and evaluate their health status} (collaboration, motivation understanding, lie evaluation), and so on. While each of these functions benefits from ToM, they are quite different expressions of what seems to be coined as the same phenomenon. This is because, in actuality, Human ToM is the result of a multitude of complex processes.

\subsection{ToM Modularity in Robotics}

While our example focuses on a future application with high requirements, the treatment of ToM as modular is abundant in the current robotics literature. Examples include Peppers being `equipped with a simple ToM module' \cite{kirtay2021trust}, introducing a ToM Manager module to deal with the representation of other's mental states \cite{devin2016implemented}, or representing ToM through a human-robot response planner \cite{gorur2017toward}. Many papers use the tri-modular `Leslie-model' \cite{leslie1994tomm}, translated to the field Robotics in Scassellati's foundational work `Theory of mind for a humanoid robot' \cite{scassellati2002theory}. This work presents a model that combines a `movement/mechanical' processor, with two ToM-like systems -- one system dealing with intents and goals of agents (i.e. actions), and one system dealing with attitudes and beliefs of agents, somewhat reminiscent of BDI, but as an extra system rather than as an integral part of the machine. This happens in for example `Theory of mind improves human's trust in an iterative human-robot game' \cite{ruocco2021theory}, where it is shown that adding processes described as part of human ToM (i.e. the attitudes/beliefs processor), positively corresponds to how trustworthy a robot comes across.\\
\\
The approaches presented by Scassellati, Ruocco, Kirtay, Devin and G\"or\"ur have one thing in common: They emulate an interpretation of beliefs, desires, intentions, and so on, but they do it through a central executive unit specifically for ToM, implying this executive unit is the representation of ToM as a whole. This sells short the complexity of the phenomenon. While the reasoning capabilities are enough for the problems set out in the papers, they moreso capture a reasoning ability about specific BDI-components and problems, into one `theory of mind'.

\subsection{ToM Modularity in Other Fields}

Research in developmental psychology has taught us that basic skills in ToM, such as the ability to recognise that others may have false beliefs, develop during childhood and early adolescence \cite{wimmer1983beliefs,wellman2004scaling,carpendale2004constructing}. The exact mechanisms that control how senses such as false belief recognition develop is unclear, but literature in developmental psychology theorises that ToM itself is influenced by at least two mechanisms: A mechanism driving subconscious snap decisions, and a mechanism that is used for elaborate, explicit, planning \cite{apperly2009humans}\footnote{This work is very reminiscent of earlier work on biases in human decision making, see \cite{tversky1974judgment,kahneman2011thinking}}. Humans make explicit representations of the mental states of others (System II), but even when unprompted and actively inhibited, humans still consider the perspective of the other (System I) \cite{samson2010seeing}. The literature also refers to this as two-systems theory, but it is argued there might be more systems that similarly emerge from ToM-style decision-making.\\
\\
We also know from the field of linguistics that ToM is closely tied to the development of language \cite{astington1999longitudinal,ruffman2002relation,de2014role, van2021modelling}. This effect is strong enough to apply bidirectionally: Experience with ToM-style reasoning also improves one's linguistic abilities \cite{slade2005language}. Reversely, being deprived of social/language input because of, for example, a hearing impairment, also appears to influence the ability to solve traditional ToM problems \cite{hall2017language}. This ties ToM to (spontaneous) conversation, painting ToM as more of a social skill, once again indicating that it is influenced by situations that draw on processes that require multiple brain regions.

\subsection{Neurological Correlates of ToM}

A direct argument against a singular ToM module comes from the many, many studies in Neuroscience that have tried to identify the brain region that is `responsible' for the human social brain and by extension, ToM. Many literature reviews find that papers in the field have varying opinions on where ToM happens in the brain, often concluding that the answer must be `in many places'\cite{carrington2009there,schurz2021toward}, or that we need a stricter definition of ToM if we wish to assign it any specific region \cite{carrington2009there,MAHY201468}.\\
\\
For example, an overview study by Carrington and Bailey, spanning 40 neurological studies, found that while both the Medial Prefrontal Cortex/Orbitofrontal Cortex (Decision-making, Expectation Management) and Superior Temporal Sulcus (Facial Recognition, Language, Voice) are `core' regions that activate for ToM-related tasks, the Temporoparietal Junction (Self-other Distinction) and Anterior- and Posterior-cingulate Cortices (Attention, Motivation, Anticipation, Learning) were also active in over half the ToM-related tasks. Many other brain regions were active for approximately 20-40\% of the tasks, indicating that solving tasks with what is considered ToM requires a large part of the brain (even if some specific regions do get more consistent activation than others).\\
\\
That said, Mahy et al.'s `How and where: Theory-of-mind in the Brain'\cite{MAHY201468} finds that different definitions in the field greatly contribute to the question of `where' ToM is located, since different definitions lead to different brain regions. They similarly conclude that both the definitions and the paradigms (e.g. evaluating cognitive vs affective states) are not precise enough. We are inclined to agree and feel these are not necessarily problems, but rather show that ToM is diverse, and that what this study encounters, characterises ToM's status as a social skill applicable in many situations. In a similar vein, Schurz et al. \cite{schurz2021toward} argue for an approach that models ToM as a multilevel construct, given that ToM tasks often reflect diverse real-world tasks that draw on different skill sets. Even for more defined tasks, such as the ToM-typical ability to recognise oneself and others' False Beliefs \cite{wimmer1983beliefs}, it is difficult to predict how other sociocognitive situations draw on the particular skill.\\
\\
The difficulties in both of the aforementioned overview studies, combined with the in-depth analysis by Carrington et al., lead us to once more conclude that ToM is a distributed process in the brain. This leaves us with an argument to see ToM as a behavioural phenomenon, underpinned by a set of mechanisms and processes that varies between tasks, contexts, and individuals, instead of a single modular function.

\subsection{Conclusion: Forgoing The Artificial Module}

This does not mean that specific implementations of ToM are a no-go. What we mainly aim to do here is create awareness. Revisiting the example of the healthcare robot, we see through the evidence in the neurosciences that humans use different parts of the brain for the robot's functions we have labelled as ToM. If we wish to reflect this in a modular fashion, the robot would have to rely on different `modules' to achieve this result in a human-like fashion. It would mean building a module for each behaviour that is functionally distinct enough to qualify as a unique expression of ToM (e.g., a `lying-recognition module', an `other-movement-reasoning' module, and so on).\\
\\
This said, we may not \emph{need} a specific (set of) module(s) labelled as `ToM'. The solution here can lie in treating ToM as a result of social skills coming together, without any regions being specifically built to represent a general or specialised ToM skill. ToM is diverse, and treating it as socially emergent might better resemble what humans do than expecting one module to be able to fulfil all of these functionalities at the same time.\\
\\
This solution does come with its own difficulties: Knowing when to address a specific aspect of ToM. Always monitoring someone's beliefs, desires and intentions is doable, but knowing how to use them (which is also ToM) is tricky. This is a discussion we consider out-of-scope, as it comes with its own set of new challenges.

\section{Every Social Interaction Requires (Advanced) ToM}\label{misc2}

Imagine you are in a hot office room. The air-conditioning is broken, you are hard at work, fully focused, and you are personally not too bothered by the heat. Your colleague at the desk to the right of you asks: ``Could you open the window for me, please?''. You do acknowledge that it's hot, and snapping out of your focus, you move yourself towards the window handle automatically, as if out of social convention -- \textit{It's hot in here, so one opens the window, it's the right thing to do}, not a single second thought. It is then that you realise that your office mate is closer to the window than you are and you have to \emph{pass} them to open the window. What was social convention and nicety before is now more of a social puzzle. \textit{Why did they ask this of me? Are they too lazy to get up themselves? Does our company policy say something about opening windows, and do they know something about this that I do not?} It is likely that we only \emph{now} have started using ToM. Initially, we only only ran the script in our brain to fulfil the favour.\\
\\
To clarify our example from a scientific point of view: There is ample evidence that we humans use scripts and heuristics in many of our daily interactions \cite{schank2013scripts,meng2008social,taylor2023reading}. We run in an almost automatic fashion, using mental shortcuts wherever we go, until it is necessary to take a more active look at the situation. For ToM specifically, this dynamic of humans being `lazy mindreaders' by default has been described by \cite{vanduijn2016}, who argues that we usually rely on flow and social synergy, until a situation demands that we `scale up' our processing efforts. This implies that we continuously (subconsciously) monitor for `hitches' that the default script/heuristic may run into, but are not always in `full-on reasoning mode'. This is understandable from a cognitive resources perspective: Using advanced ToM right-away is a major resource drain as it is quite costly to use \cite{lin2010reflexively,lewis2017higher}. It is also quite prone to error as this analysis quickly falls into overthinking perspectives on what the other party thinks about \textit{you} in turn, as it is very difficult for humans to visualise perspectives recursively without making any mistakes \cite{wilson2023recursive}.\\
\\
This leaves a major question, both if we want to understand humans, but also if we want to build systems that do: `What are the origins and nature of all these scripts that humans use?' Clark argues that starting our reasoning from what he refers to as `Common Ground' \cite{clark2006} already resolves a lot of complexity that other researchers have assumed or identified \cite{ogrady2015}. Quite similarly, it has been shown in AI research that one can reason effectively using experience-based patterns in a human-like fashion \cite{mcilroy2020aligning}, and that in (at least) the medical domain, abstracting single beliefs into higher-level abstractions is an efficient method of dealing with complex domains \cite{erdogan2022abstracting}. In the next section, we explore additional behaviours that come across as ToM on a behavioural level, but do not make use of ToM on a mechanistic level (as explained in Section \ref{definitions}).

\subsection{ToM: A Collaborative and Competitive Skill?}

Successful cooperation and collaboration are often designated as resulting from the use of Human-ToM. This thought is especially prevalent in the field of evolutionary anthropology, where it is often assumed that ToM has played a major part in human evolution and its eventual development towards a highly social species \cite{stone2006theory, tomasello2013origins}. Much of this is expressed in the Social Brain Hypothesis, the notion that social complexity was an important driver behind the evolution of intelligence in primates and various species across other orders \cite{byrne1996machiavellian, dunbar2007understanding}. It has been shown that various aspects of species' social life, such as group size or mating style, are reflected in their brain size and brain organisation, as well as in social-behavioural repertoire.\\
\\
Indicators of this have also been shown through agent modelling. ToM has been shown to be a successful strategy when modelling evolution of social behaviour, resembling human behaviour better than making fully rational decisions in the Incremental Centipede Game \cite{lenaerts2024evolution}, and research into the negotiation game Coloured Trails has shown that ToM is a skill that grows in benefit as both the environment and resource dilemmas become more complicated \cite{de2022higher}.\\
\\
We do not wish to argue with or contest these findings: ToM is provably useful in both collaborative and competitive settings, and even more so in combinations of the two \cite{verbrugge-mixed}. Yet we do argue that ToM is neither always required nor used by default. A first example of this comes from a model that plays the aforementioned Coloured Trails without making use of explicit ToM \cite{ficici2007modeling}. This model trained on human gameplay data is able to perform with similar effectiveness to the original ToM coloured trails model \cite{de2014effectiveness}, using only statistical approximation. This performance holds up for multiple ToM levels (it was not trained beyond ToM level-2 due to the computational complexity). This is an initial indication that if a system is exposed to enough experience with a situation (e.g., the human gameplay data), it is able to perform similar to ToM.\\
\\
Additional examples of this exist when we analyse more traditional boardgames. Both Chess \cite{campbell2002deep} and Go \cite{silver2016mastering, silver2017mastering} have been solved through the use of statistics and emulated experience, even though both games have been argued to be heavily (ToM)-reasoning dependent. The models were purely trained by extensive self-play, turning the problem into one of search optimisation. What is more, similar to the non-ToM-human-play Coloured Trails model, models for Chess that are able to capture a human-style of play without using explicit ToM \textit{do} exist (such as `Maia') \cite{mcilroy2020aligning}. These models rely neither on ToM nor on pure probabilistic reasoning. The major benefit of this system is that it can play chess much more like a human would, at the specific chess ranking (ELO) the human corresponds to. Its abstractions are able to predict the human player's next move in approximately half of the cases for positions when there are multiple `sensible' options. Effectively, the system is able to capture human-like competitive behaviour in Chess without the need for explicit reflection on the competitor's mental state, instead relying on pre-trained human data applied to the state space (the game board).\\
\\
Research has also shown that it is possible to emulate computational ToM by abstracting single beliefs into higher-level concepts \cite{erdogan2022abstracting}. Using the value of pre-established roles, social values, and social norms, can short-cut a decision that would otherwise require advanced reflection with high accuracy, at least in the medical domain. It has similarly been shown that in cooperative counting tasks, the use of a previously established strategies and synergy is sufficient to find good solutions without the need of continued explicit reflection on one's partner, provided the rules of the game or the collaboration partner do not change \cite{meulen2024common}. In this particular case, establishing this strategy did require a mental model and explicit reflection on one's partner to reach this point -- the point is not that ToM is redundant, but that it is no longer necessary after a certain level of familiarity (and perhaps implied trust) has been established.

\subsection{Revisiting the Human Perspective} \label{misc2-humper}

Knowing ToM can occasionally be replaced by different strategies, that use experience or other reasoning short-cuts, gives us a perspective on how the cost \cite{lewis2017higher} and inaccuracy \cite{wilson2023recursive} problems of ToM can be dealt with in practice. However, the knowledge that these alternatives exist, does not answer when humans do, and when they do not use ToM, especially not when they use the intensive `higher-level' ToM so famous for this costliness.\\
\\
So, how advanced does ToM need to be, both for human-human collaboration, and for human-AI collaboration? To answer this question, it is important to evaluate its value. Revisiting the Coloured Trails negotiation simulation, we are led to conclude that ToM's benefits drop off after three levels of recursion, both in experimental agent-agent settings, but also when an agent negotiates with a human \cite{de2014agent}\footnote{In negotiation, the 3rd level of recursive ToM would be: Taking into account that your partner will account for you accounting for them.}. The level that humans realistically take seems to be slightly higher (level-3, level-4, presumably even higher) in tasks that are perceived as less mentally taxing, as indicated by research that solves the Mod-game (a numerical variant of Rock-Paper-Scissors with proximity-based point-scaling) \cite{frey2013cyclic} using the same recursive ToM setup \cite{veltman2019training}. In other words: depending on the task, there is a benefit and realistic usage drop-off, even if this drop-off depends on the complexity of the task.\\
\\
Looking at situations with more complex pay-off structures and higher mental effort, the same recursive reasoning setup reveals that humans sometimes already have trouble with using second-order reasoning \cite{verbrugge2018stepwise}. They can, however, become better in ToM-reasoning through explicit step-wise training. This also holds for higher orders of ToM \cite{valle2015theory}. A key component to ToM reasoning seems to be the awareness that ToM-like reasoning is required to begin with, which is something humans still struggle with in adulthood \cite{keysar2003limits}. Experience with the topic helps in these cases \cite{samson2010seeing,apperly2013can}. How deeply human ToM reflections go, however, still differs per person \cite{stiller2007perspective}, and per skill level \cite{apperly2009humans}, which is yet another indicator that even in adulthood, one's experience with the situation is key to their problem-solving aptness.\\
\\
There is a crucial realisation here: Humans often do not apply ToM automatically unless prompted, and experience and familiarity with the topic weigh heavily into how well this ToM is applied once prompted. It would be unwise to overestimate the capacity of a human collaboration partner, both when we are a human, and when we are an AI system. The behaviour from the human is probably better explained using a short-cut strategy, or a relatively low level of ToM (level-one ToM, only reflecting on our partner's thoughts on the situation, meaning level-two would be sufficient from either side in hopes of this leading to a successful synergy).

\subsection{Conclusion: Working Smarter}

The use of higher ToM levels, going beyond a first-order reflection on one's partner's perspective, will generally require more effort, and is often unnecessary. If the situation is not very dynamic, using too high of a ToM level might result in a non-human-like algorithm, that overestimates a human partner's reasoning methods. In human-human interaction, many collaborations rely on pre-established or predictable patterns. We recommend a similar approach in AI systems: Rely on scripts and heuristics by default, and only defer to active ToM reasoning only when needed.\\ 
\\
In short, it is not necessary for \emph{every} ToM system to be `as good as possible'. Only if we need an AI system that is an expert on the situation, or has to be in an overseer position that specifically works with (human) experts on the domain, we may need higher levels of advanced, explicit, ToM. 

\section{All ToM Is The Same}

Humans have a tendency to somewhat personalise their use of ToM based on their experiences with a person: Knowing someone, or even having a superficial first impression, impacts how we, as humans, think about one another. If a person comes across as quite snobbish, we will have a different impression of them as when they come across as an enthused savant, and our reasoning about these two individuals will be different. In the case of the snob, we may feel like they have met people like this before: the know-it-all, who believes their tastes superior to someone else's (and looks up to very specific individuals). They will know a lot about a specific niche subject, be very expressive, sometimes even aggressive, about their knowledge on that topic, and might exaggerate that knowledge to a related-but-slightly-broader domain. We might be wrong about these assumptions, and might even be aware that we are biased about this `snob' based on previous experiences, but we will still rely on the bias quite often (referring back to Section \ref{misc2-humper}: Biases and stereotypes are an example of a shortcut or heuristic in the social domain).\\
\\
These stereotypes are useful and effective, but intuitively, they harm collaboration on a deeper level. We `know' the snob thinks differently than us about some things (even if we may be wrong about some), and so far this is useful for collaboration. What we do not know, is how the snob really thinks about \emph{us}, as we cannot draw an accurate picture of what biases the snob will have about us (only an approximation) unless they ask and get a truthful answer. Perhaps the snob thinks of us as very interested in what they have to say, estimating their trust in us to be at a far higher than it actually is (after all, the snob is presumably quite arrogant). What is an optimal choice for one individual, is not for the other, especially from a social rather than an economic point of view. Different people have different models of one another, and those models are sometimes biased to a degree that hampers collaboration. We need a `snob' translation manual to fully do the `correct' mentalisation. Knowing this, why try to capture ToM in a one-size-fits-all `generalisable' model? In this section, we elaborate on the tested differences in ToM across entities.

\subsection{Human-Human Interaction} \label{humanhuman}

Intuitively, it is easy to see why robots, all sorts of animals\footnote{Whether many types of animal thought can be classified as ToM to begin with is very much an open empirical question, with many shades of doubt \cite{call2008does,van2012corvid,barrett2014dolphins}}, and humans think differently. And that the snob's mentalisation functions differently than the non-snob. Maybe less intuitively, it needs to be said that ToM differs \emph{between} `categories' of humans as well, regardless of their specific personality (although this also matters). Across human populations, there are a few factors that greatly influence their `use' of ToM.\\
\\
A strong characteristic of the accuracy of one's use of ToM is how similar the other individual is. An example drawn directly from ToM literature in developmental psychology is research into autistic individuals, and the long discussions on the  supposed lack of ToM in these individuals. The conclusion on their problems with ToM is heavily flawed, see for example \cite{gernsbacher2019empirical}. In what is referred to as the `Double Empathy problem' \cite{milton2012ontological}, it is argued instead that autistic individuals have less trouble understanding (and thus mentalising) other autistic individuals, and that the same holds between allistic/neurotypical individuals, but that both `categories' sometimes have trouble making predictions about each other's behaviour and mental states and motivations\footnote{This is in practice a fair bit more nuanced, since neurodivergence is a spectrum, but similarity on the spectrum is presumed to aid with mentalising the other}.\\
\\
The `individual similarity' principle also holds for other groups that are on some level generalisable. For example, culture is of significant influence when it comes to one's mentalisation tendencies. Seminal work by Wellman and Liu has categorised the development of ToM into an order by which children tend learn ToM-related skills \cite{wellman2004scaling}: they first learn to understand one may [i] desire different things, then that 
 [ii] one may believe different things, [iii] one may have different knowledge, [iv] meaning one may have false beliefs, which can later be 
 [v] explicitly represented false beliefs, and one can then [vi] have emotions to do with these beliefs after which they [vii] can lie about their emotions.\\
 \\
 This order seemed to be pretty clear-cut. However, later research showed this order is different depending on what the local cultural values about some of these concepts are. The Wellman-Liu ordering more strongly applies to cultures that generally value beliefs over knowledge (which can probably be linked to individualism). In more colloquialism-based cultures, that generally value knowledge over beliefs, children instead learn to recognise different ToM-states on knowledge before they learn to recognise different ToM-states on beliefs, also influencing the timetable of false-belief recognition (e.g. Iranian \cite{shahaeian2011culture,shahaeian2014iranian} and Chinese children \cite{wellman2006scaling,liu2008theory}).
\\
\\
The power that culture has over shaping one's mind is especially present during development \cite{lev1979mind,tomasello2010gap}. The differences in development `priorities' between colloquialism- and individualism-driven cultures implies that the exposure children have to these ToM-aspects varies significantly. After all, more strongly weighing one aspect of ToM over the other, actively influences the priorities in perspective-taking the culture transfers to its next generation\footnote{Note that there is no overall difference in the `ability to understand aspects of ToM to begin with': in the end, children who have enough social (and language) exposure develop each of the listed skills, although they may lean toward mind-reading based on one of the listed skills over the other \cite{liu2008theory}}.\\
\\
This is not the only aspect where the culture one has grown up in is influential: It also affects how humans mentalise others on a less ingrained scale. When stories use protagonists and objects familiar to a human's specific culture, that human will more often (and more speedily) make the correct ToM assertion, even if the situation described is not inherently cultural (f.e. a complex false belief task) \cite{perez2016cultural}. Additionally, growing up with input from two or more cultures, and even self-evaluated openness to other cultures, both aid in mind reading people from other cultures in general \cite{kim2023mindreading}. Note that this study has also shown that, true to any discussion on specialisation versus generalisation, under the same culture, mono-culturalists seem to be better at mind-reading people from that specific culture than multiculturalists.\\ 
\\
Knowing that \textit{on average} similarity and openness to cross-cultural experiences are both beneficial to successful mind-reading efforts provides us with a solution on improving specific human-human interaction: awareness. Spending active and conscious effort into making the humans in a collaboration aware of the aspects that matter to their mind-reading, and asking them to reflect on these matters with the right materials, should nudge both parties in the interaction towards more productive mind-reading. 

\subsection{Human-AI Interaction}

Of course, the concepts that apply to human-human interaction, presumably also apply to some degree in human-AI interaction. What we add on top of this, is that AI systems also lack a more general `human' experience. So far, what we have seen, is that common ground between entities \cite{clark2006} is a major contributor to how well they understand each other. Humans share a same biology, have the same sensors they use to interact with the world\footnote{How these sensors make sense of the world can still be very different!}, and know that all other humans have this as well. There may be many mental differences, but they are grounded by needing sustenance, and by their beliefs, desires and intentions. The divergence of these grounds starts at a different level (and actively happens through what we call `grounding' \cite{clark2015common,geurts2018convention}.\\
\\
This is not the story of the AI system. The AI system does not have the same sensors as a human, has a very different `brain', does not need food or water... and as such misses an intricate ground with humans \cite{brooks2018what}. It is the designer's task to allow a human and an AI system to bridge this gap. When talking about specific states of an object, of knowledge, of a belief, etc., the human and the AI need to ground whether they are on the same page (sadly more often than a human and a human, as the gap is larger). In specialistic tasks, this is doable, as the domain is limited and the possibility to add knowledge to a pre-defined domain can be added into a system: Things may not be truly human, but the human and AI can both, in their own way, reason about the knowledge/world states that they now both have. For more general AI, a solution lies further on the horizon.\\
\\
It is very important for the human to realise the difference in ground that they and the system have. As we know from human-animal interaction, it is very easy for humans to attribute ToM to animals, as communicative intention with the entity inherently draws out attribution of an assignment of mental states \cite{airenti2018development}. This holds especially for pet owners \cite{eddy1993attribution}, but is also true in general. Exposure to human-like animals in fiction strongly raises the human tendency to apply ToM to animals in real life, because of the `blueprint' impressions these fictionalised animals leave that are then applied to real-world animals \cite{grasso2020anthropomorphized}. This is despite evidence to the contrary \footnote{Note that this on its own is an indicator that humans prefer quick strategies over the use of rationalising their way through evidence, in reference to the previous misconception}.\\
\\
Returning to human-AI interaction: This problem is similar when we look at (especially) human-robot interaction. A robot using human speech patterns is perceived as `more human' \cite{eyssel2012if}, looking like a human increases a robot's perceived intelligence  \cite{fink2012anthropomorphism}, and a robot expressing human-like attitudes (i.e. express actions one would also evaluate during ToM-like reasoning) is considered more friendly \cite{bernier2010similarity}. In practice, all of these aspects are but part of a whole that combines into a concept of anthropomorphisation.\\
\\
Having established that the differences between a human and an AI system are vastly significant for how they deal with beliefs, desires and intentions, this is a problem. Anthropomorphism can be a harmful bias for human-AI -- one that can be worked on to overcome -- but it becomes more problematic when it becomes a practice that is actively leveraged. If a system is anthropomorphised on purpose, without any active mechanisms that allow it to genuinely ground with its human interactants, it will result in a lot of ToM misconstrual.

\subsection{Conclusion: Finding Common Ground}

What do these insights mean for AI \& CS developers and researchers? A preliminary advice: Do not fake it. It is not enough to \emph{seem} human-like: The system should not trigger the user to make unfounded assumptions about the AI system's human-likeness, even if the aforementioned anthropomorphism raises trust in the system \cite{wang2021towards,placani2024anthropomorphism}. Be open about what the system can and cannot do. Try to create common ground with the system where possible, so stay inquisitive: Collect information to adjust the mental map of both the human user and the AI system.\\
\\
Ultimately, we suggest two solutions. The first solution is that humans need to invest in building up a `theory of AI mind', i.e., learn how the specific AI system reasons, functions, and acts in the world. Developers need to explain to the user how the specific system maps its beliefs to behaviour, and how its sensors perceive the world. On the flip-side, the system needs to clearly communicate its intentions to the human user and explain how it interprets what the human is doing (again, trying to find common ground). Our shared responsibility as developers is making sure that these interpretations contain a conceptualisation that can be translated to correct actions, to position the AI system as Dennett refers to it, an intentional one \cite{dennett1971intentional}. Alternatively, the system needs to be designed in a human-like way, so that humans can apply what they know about interaction with humans to interacting with this AI system. This is, however, a far trickier solution, since it requires a thorough understanding of emulating the human brain.

\section{Current AI Systems Already Have ToM}

Computational models of ToM have a long tradition in agent-based modelling, including recursive, Bayesian, and neural frameworks \cite{de2017negotiating, baker2011bayesian,jin2024mmtom,pmlr-v80-rabinowitz18a}. Each of these frameworks has their own individual claim to ToM, but there are differences how and for what purpose this capacity is modelled. In this section, we discuss these properties, and the misconceptions relating to these properties. Additionally, we elaborate on the perceived ToM behind a specific implementation of a neural framework: ToM with respect to large language models (LLMs).\\
\\
Recursive ToM-based models are usually rooted in reasoning that is based on utility, epistemic logic, doxastic logic, and combinations thereof. Their deterministic nature makes them quite useful for modelling exact knowledge and belief states, in ways that also enable a model to rationally reason about both the knowledge it has and the knowledge it has yet to acquire. This makes these models perfect for ToM-scenarios that rely on the resolution of mental states to the nth level (ToM level-1, level-2, level-3, etc.) in an explainable, traceable, way, enabling them to take a clearly defined (counter)action based on their interaction partner's perspective (BDIs). These models do require a mechanism that determines the relevance and importance of these BDIs. This mechanism is often represented by economic rationality' the most optimal action from the perspective of their partner (depending on the ToM level), which often does not match with reality. We have already questioned this notion in ``Every Social Interaction Requires (Advanced) ToM'' (Section \ref{misc2}), but would like to reiterate that humans that use short-cuts and lazy reasoning are often not rational. This makes these systems likely to reason erroneously if applied in real-world social settings, even if it makes them perfect for resource negotiation and strategy simulations (and in a collaborative setting, a far better reasoner when dealing with deliberate experts than when dealing with novices!).\\
\\
Bayesian models, in turn, are able to better incorporate non-rational actions as they can model hidden behavioural properties when enough data is available. They do only function realistically (with a high performance rate) when the priors are right, leveraging a lot of real-world information that may not be practically available, as information existing in the world, through statistical patterns, does not automatically imply this information is also retrievable by the designer. In terms of realism, although the underlying model might \textit{map} human ToM in a seemingly realistic fashion, we know from experience that humans prefer shortcuts and are not very probabilistically driven, which one needs to take in constant account when modelling their agents in a Bayesian fashion.\\
\\
We will focus on Machine Learning-based Systems in more detail, due to the attention Large Language Models have drawn with respect to specifically ToM, driving a need to analyse a potential misconception.

\subsection{``Machine Learned Systems Have A Theory of Mind"}

Recent developments in neural frameworks have seen rapid developments in the human-likeness of AI: Generative pre-trained transformer models, forming the basis for LLMs and large multimodal models (LMMs), have very convincing linguistic capabilities, especially when looking at their most recent iterations\footnote{Examples of commercial, closed-source models are OpenAI's ChatGPT \cite{brown2020language, achiam2023gpt} and Google's Gemini \cite{geminiteam2024geminifamilyhighlycapable}; open-source variants are e.g. the OLMo model family \cite{groeneveld2024olmoacceleratingsciencelanguage}.}. Their success in several interactive contexts has sparked debate over whether a form of ToM may have emerged in such models. This is not an unreasonable claim, given the intricate ties between ToM and language in human development and evolution \cite{vanduijn2016} -- and thus warrants an in-depth look.\\
\\
LLMs beyond a certain size and level of fine-tuning pass traditional false-belief tests \cite{kosinski2024evaluating,Strachan2024NHB}, i.e. tests that have been developed to assess ToM competence in human children and specific populations \cite{baron1985,barone2019infants}. However, such performance was mostly attributed to ample presence of the benchmarked ToM tests in the training data, meaning that superficial task recognition was sufficient to give the right answer. When tests were adapted to avoid this, LLM performance was shown to drop \cite{ullman2023large,shapira2023clever} or in need of nuance \cite{van-duijn-etal-2023-theory}.\\ 
\\
Despite our own scepticism whether performance and standardised tests really indicate ToM abilities, we do feel that general rebuttals sell the story a bit short. LLMs were neither designed nor trained specifically to perform ToM tasks, and we have seen from studies in developmental psychology that some aspects of ToM \emph{can} indeed be an emergent property of language acquisition \cite{de2014role,milligan2007language}.\\
\\
Following the initial results and debate, ToM benchmarks were introduced \cite{kim-etal-2023-fantom, chen2024tombenchbenchmarkingtheorymind, wang2024tmgbenchsystematicgamebenchmark}, comparisons were made against human (child) scores \cite{van-duijn-etal-2023-theory, Strachan2024NHB}, other modalities were integrated \cite{razothesis2024, strachan2024gpt4oreadsmindeyes}, integrations with older model architectures were explored \cite{jin2024mmtom}, and theoretical reflection was added \cite{goldstein2024doeschatgptmind}. Resulting from this literature, we briefly reflect on three additional aspects indicating that the abilities of LLMs should to be taken seriously, but are far from being there.

\subsubsection{Correlation with Real-World Social Ability}

The observed ability of LLMs to score well on standardised ToM tests may not correlate with real-world social abilities \cite{van-duijn-etal-2023-theory}. In principle this is an inherited flaw of how humans are often evaluated in their ability to solve ToM-related problems, as these evaluations, similar to LLMs, rely heavily on linguistical ability \cite{barone2019infants}. However, for humans, a large body of work associates performance on standardised ToM tests with various landmarks in children's socio-cognitive development \cite{beaudoin2020systematic}. These landmarks go beyond the linguistic abilities that these ToM tests are biased towards, meaning that ToM tests can reasonably be argued to be an effective means of evaluating the human ability to mentalise: No comparable landmarks exist for LLMs.\\
\\
For \emph{LLMs}, it is thus an open empirical question how well test scores generalise to their social competencies in \textit{actual} interactions with humans. We have no such corresponding socio-cognitive data for LLMs, as LLMs have no `lived' experience, only knowledge driven by their text- and static image-heavy training corpora \cite{vandijk2023largelanguagemodelsneed}. As a consequence, while LLMs \emph{may} be able to generalize beyond statistical pattern prediction, the knowledge of LLMs is very differently (and probably much less firmly) grounded compared to humans's (see also the discussion on Misconception 3, ``All ToM is the Same'').

\subsubsection{Perspective Grounding}

There is an important distinction between exposure-based \textit{third-person} and experience-based \textit{first-person} social reasoning. This is not properly captured by traditional benchmarks. Most benchmark ToM tests, also those in lab settings, take an `observer perspective'. These tests may provide LLMs with an `unfair' advantage given a training set with ample descriptions of social life from, e.g., literary fiction or online fora with people sharing their experiences in everyday life (all top-down provided information, exposure). However, once more: Answering questions about social situations from an outside perspective differs greatly from actually engaging \textit{in} such situations.\\
\\
Specific explorations into this topic have shown that, indeed, when an LLM is forced to take individual perspectives through dialogue (i.e., a first-person perspective), its performance sharply declines \cite{kim-etal-2023-fantom}. In the first-person-dialogue case, models were quite likely to incorporate characters that were unaware of the presented information into their chain of reasoning. Additionally, presenting more than just directly relevant information, involving characters who do not have mental states about the subject, with the systems often unable to identify relevant concepts from irrelevant ones. Needless to say, this differs significantly from the interaction-driven ToM that humans often experience, where keeping track of information relevance and information availability are key to human social skill.\\
\\
Similarly, Hou et al. found that using first-person perspectives severely hampers LLM performance in even the currently most advanced LLMs, such as GPT 4, Claude 3.5 and Llama 3.8 \cite{hou2024enteringrealsocialworld}. In contrast, converting the same (textual) social situation to one presented in a 3rd person, more narrative, style, majorly bumped up the performance of the systems.

\subsubsection{Scale over Substance}

We know that model characteristics, test type, and test approach influence performance. It is no coincidence that more recent models, fed with more data than ever before, outperform the older models. The claims that ToM performance reach teen levels \cite{webb2023emergent,street2024llms} can sometimes even be refuted in-paper by controlling for the associative reasoning that LLMs have gotten quite good at due to amount of data they are trained on: Forcing a relational (logical) reasoning style instead of an associative one shows major differences in ToM performance (1-2 years vs early teens) \cite{stevenson2023large}.\\ 
\\ 
Additionally, fine-tuning and prompting approaches boost scores on ToM-standardised tests \cite{ma2023tomchallenges, moghaddam2023boosting, huang2024notioncomplexitytheorymind}. The study by Moghaddam and Honey reveals that ToM tests that LLMs do still struggle with, are significantly better executed when human feedback is added to the system through reinforcement learning. This casts doubt on the idea that the exposure-based techniques that LLMs are driven by, is enough to capture human ToM. In fact, the additional training through human feedback was beneficial, but modern LLMs were still unable to solve the complex second-order ToM tasks at hand.

\subsection{Conclusion: Nuance Needed in the Debate over ToM in LLMs}

What we learn from LLMs struggling to successfully adopt a first-person ToM perspective is the importance of real-life experience within social settings. This is also our advice for future benchmarks: Evaluate your systems through social settings, \textit{in practice}, rather than through benchmarks adapted from tests developed for humans. This is understandably a big ask, with a potentially high engineering effort, but to us, it does feel like the only way to evaluate the aspects of ToM that are considered to be informed attributes in humans. Expose the system to real-life human behaviour: ToM as a social skill should not be lab-bound, especially knowing how the framing of a setting can already determine whether even a human uses ToM or not, and whether or not this is successful. If we do not evaluate a system's ability to reason beyond limited benchmarks, we may not end up testing how truly socially dynamic such a system is. ToM is a skill that deals with the unexpected, applied in specific settings: Research into emulating ToM for our AI systems should ensure the situation is \textit{actually} unexpected.

\section{Discussion and Concluding Remarks}

We have defined ToM as the ability to make sense of someone's behaviour and attitudes towards the world by reasoning from their perspective, in terms of their beliefs, desires, intentions, motivation, and so on. AI systems can make estimations about beliefs, desires, and intentions, in specific situations, using specific techniques. Calculated problems, such as a resource negotiation, a false belief puzzle, a prediction on what direction someone will walk into, all benefit from ToM, but are specialised instances of a far bigger whole. Generalisability beyond these instances is challenging for even the most capable AI systems. Yet we have discussed that ToM might not be as generalisable of a skill in humans either, often acting as a function of how experienced a human seems to be with the specifics of the situation, which is linked to individual and socio-cultural differences.\\
\\
Humans tackle the situation head-on and learn on the go, modelling their collaborative partners and not-so-collaborative competitors in the new and specific situation as they learn new things about them. Similarly, online and dynamic learning feels like a feasible approach for many AI systems as well, even if this is a demanding process. It is the human ability to learn that creates the apparent flexibility. After all, `What would I do, if I were in their shoes?' is a valid question when the situation calls for a suspicion of one's motives, but is effortful and might require a manual to translate between cultures or different entities altogether. This manual needs to be either pre-delivered to aid with the translation, or the entities need to be similar enough to ourselves to not require such a translation to begin with. What often leads to misunderstandings or worse, is to simply assume this similarity.\\
\\
We end by pointing to an alternative approach: the notion of Hybrid Intelligence \cite{dellermann2019hybrid,akata2020research}. Humans and AI Systems each have their separate strengths. While keeping the discussed misconceptions in mind, the strengths of both can be leveraged. AI Systems have a good memory, fast retrieval, can quickly make difficult calculations, and are very good at finding patterns in data. Humans are better at evaluating the pragmatic value of these memories and patterns, resolving ambiguity in communication, and at dealing with social situations and complexity in general. These skills can be made complementary, as long as a cooperative setting is designed for. Revisiting our Healthcare Robot: The AI system driving the robot might be better at detecting when something is off, noticing a (minor) change in behavioural or other patterns. The human side of this duo would be able to collaborate with the robot and work out the nuanced social dynamics of the situation. For example, the robot could determine the relevant questions to ask a patient, whereas the human would be able to formulate them adequately and adapt to the direct responses they evoke. In the end, through such forms of collaboration, humans and AI might even learn something from each other and improve their ToM skills on both ends.

\newpage
\bibliographystyle{ieeetr}
\bibliography{biblioToMAICS}

\end{document}